\documentclass[11pt]{article}
\usepackage{amssymb,amsmath,amsfonts}
\usepackage{graphicx}
\usepackage{graphics}
\usepackage{eepic,epsfig}

=1.60cm

\begin{document}

\title{Casimir densities for a boundary in Robertson-Walker spacetime}
\author{A. A. Saharian$^{1}$\thanks{%
E-mail: saharian@ictp.it}\, and M. R. Setare$^{2}$ \thanks{%
E-mail: rezakord@ipm.ir} \\
\\
\textit{$^1$Department of Physics, Yerevan State University,}\\
\textit{1 Alex Manoogian Street, 0025 Yerevan, Armenia}\vspace{0.3cm}\\
\textit{$^2$Department of Science, Payame Noor University, Bijar, Iran} \\
}
\maketitle

\begin{abstract}
For scalar and electromagnetic fields we evaluate the vacuum expectation
value of the energy-momentum tensor induced by a curved boundary in the
Robertson--Walker spacetime with negative spatial curvature. In order to
generate the vacuum densities we use the conformal relation between the
Robertson--Walker and Rindler spacetimes and the corresponding results for a
plate moving by uniform proper acceleration through the Fulling--Rindler
vacuum. For the general case of the scale factor the vacuum energy-momentum
tensor is presented as the sum of the boundary free and boundary induced
parts.
\end{abstract}

\section{Introduction}

The influence of boundaries on the vacuum state of a quantum field leads to
interesting physical consequences. Well known example is the Casimir effect
\cite{Most97,Milt02,Bord09,Eliz94}, when the modification of the zero-point
fluctuations spectrum by the presence of boundaries induces vacuum forces
acting on the boundaries. It may have important implications on all scales,
from cosmological to subnuclear. The particular features of the resulting
vacuum forces depend on the nature of the quantum field, the type of
spacetime manifold, the boundary geometries and the specific boundary
conditions imposed on the field.

The Casimir effect can be viewed as a polarization of vacuum by boundary
conditions. Another type of vacuum polarization arises in the case of an
external gravitational field. In this paper, we study an exactly solvable
problem with both types of sources for the polarization. Namely, we consider
the vacuum expectation value of the energy-momentum tensor for both scalar
and electromagnetic fields induced by a curved boundary in background of
Robertson-Walker (RW) spacetime with negative spatial curvature. In order to
generate the vacuum densities we use the well known relation between the
vacuum expectation values in conformally related problems (see, for
instance, \cite{Birrell}) and the corresponding results for an infinite
plane boundary moving with uniform acceleration through the Fulling--Rindler
vacuum.

The latter problem for conformally coupled Dirichlet and Neumann massless
scalar fields and for the electromagnetic field in four dimensional Rindler
spacetime was considered by Candelas and Deutsch \cite{CandD}. These authors
consider the region of the right Rindler wedge to the right of the barrier.
In \cite{Saha02}, we have investigated the Wightman function and the vacuum
expectation values of the energy--momentum tensor for a massive scalar field
with general curvature coupling parameter, satisfying the Robin boundary
condition on the infinite plane in an arbitrary number of spacetime
dimensions and for the electromagnetic field. We have considered both
regions, including the one between the barrier and Rindler horizon. The
corresponding surface densities induced on the plate have been considered in
\cite{Saha04c}. The vacuum expectation values of the energy--momentum
tensors for scalar and electromagnetic fields for the geometry of two
parallel plates moving by uniform acceleration are investigated in \cite%
{Avag02}. In particular, the vacuum forces acting on the
boundaries are evaluated. In \cite{Saha04} the Casimir energy is
evaluated for massless scalar fields under Dirichlet or Neumann
boundary conditions, and for the electromagnetic field with
perfect conductor boundary conditions on one and two infinite
parallel plates moving by uniform proper acceleration through the
Fulling--Rindler vacuum in an arbitrary number of spacetime
dimension.

A closely related problem for the evaluation of the
energy-momentum tensor of a Casimir apparatus in a weak
gravitational field recently has been considered in
\cite{Full07,Bimo08}. In particular, it has been shown that the
Casimir energy for a configuration of parallel plates gravitates
according to the equivalence principle. In \cite{Saha04b} the
conformal relation between de Sitter and Rindler spacetimes is
used to generate the vacuum expectation values of the
energy--momentum tensor for a conformally coupled scalar field in
de Sitter spacetime in the presence of a curved brane on which the
field obeys the Robin boundary condition with coordinate dependent
coefficients. The Casimir densities for spherical branes in
Rindler-like spacetimes have been investigated in \cite{Saha05}.

The organization of the present paper is as follows. In the next section we
consider the conformal relation between the problems in the RW spacetime
with negative spatial curvature and Rindler spacetime. The geometry of the
boundary is specified. In section \ref{sec:Scalar} the vacuum expectation
value of the energy-momentum tensor is investigated for a scalar filed with
Robin boundary condition. The Casimir densities in the case of the
electromagnetic field with perfect conductor boundary conditions on the
plate are discussed in section \ref{sec:Elmag}. The main results are
summarized in section \ref{sec:Conc}.

\section{Conformal relation between the problems in Robertson-Walker and
Rindler spacetimes}

\label{sec:ConfRel}

As a background geometry we shall consider the $k=-1$ RW spacetime with the
line element%
\begin{equation}
ds^{2}=g_{ik}dx^{i}dx^{k}=a^{2}(\eta )(d\eta ^{2}-\gamma
^{2}dr^{2}-r^{2}d\Omega _{D-1}^{2}),  \label{dsRW}
\end{equation}%
where $\gamma =1/\sqrt{1+r^{2}}$ and $d\Omega _{D-1}^{2}$ is the line
element on the $(D-1)$-dimensional unit sphere in Euclidean space. First of
all let us present the RW line element in the form conformally related to
the Rindler metric. With this aim we make the coordinate transformation%
\begin{equation}
x^{i}=(\eta ,r,\theta ,\theta _{2},\ldots ,\theta _{D-2},\phi )\rightarrow
x^{\prime i}=(\eta ,\xi ,\mathbf{x}^{\prime }),  \label{CoordTrans0}
\end{equation}
with $\mathbf{x}^{\prime }=(x^{\prime 2},...,x^{\prime D})$, defined by the
relations (see Ref. \cite{Birrell} for the case $D=3$)%
\begin{eqnarray}
&& \xi =\xi _{0}\Omega ,\; x^{\prime 2}=\xi _{0}r\Omega \sin \theta \cos
\theta _{2},\ldots ,\;x^{\prime D-2}=\xi _{0}r\Omega \sin \theta \sin \theta
_{2}\cdots \sin \theta _{D-3}\cos \theta _{D-2},  \notag \\
&& x^{\prime D-1} =\xi _{0}r\Omega \sin \theta \sin \theta _{2}\cdots \sin
\theta _{D-2}\cos \phi ,\;x^{\prime D}=\xi _{0}r\Omega \sin \theta \sin
\theta _{2}\cdots \sin \theta _{D-2}\sin \phi ,  \label{CordTrans}
\end{eqnarray}%
where $\xi _{0}$ is a constant with the dimension of length and we use the
notation%
\begin{equation}
\Omega =\gamma /(1-r\gamma \cos \theta ).  \label{Omega}
\end{equation}%
Under this transformation the RW line element takes the form%
\begin{equation}
ds^{2}=g_{ik}^{\prime }dx^{\prime i}dx^{\prime k}=a^{2}(\eta )\xi
^{-2}\left( \xi ^{2}d\eta ^{2}-d\xi ^{2}-d\mathbf{x}^{\prime 2}\right) .
\label{dsRW1}
\end{equation}%
In this form the RW metric is manifestly conformally related to the metric
in the Rindler spacetime with the line element $ds_{\mathrm{R}}^{2}$:%
\begin{equation}
ds^{2}=a^{2}(\eta )\xi ^{-2}ds_{\mathrm{R}}^{2},\;ds_{\mathrm{R}%
}^{2}=g_{ik}^{\mathrm{R}}dx^{\prime i}dx^{\prime k},\;g_{ik}^{\prime
}=a^{2}(\eta )\xi ^{-2}g_{ik}^{\mathrm{R}}.  \label{ConfRel}
\end{equation}

By using the standard transformation formula for the vacuum expectation
values of the energy--momentum tensor in conformally related problems (see,
for instance, \cite{Birrell}), we can generate the results for the RW
spacetime from the corresponding results in the Rindler spacetime. First we
shall consider the corresponding quantities in the coordinates $(\eta ,\xi ,%
\mathbf{x}^{\prime })$ with the line element (\ref{dsRW1}). These quantities
are found by using the transformation formula for conformally related
problems:
\begin{equation}
\langle 0_{\mathrm{RW}}|T_{i}^{k}\left[ g_{lm}^{\prime },\varphi \right] |0_{%
\mathrm{RW}}\rangle =[\xi /a(\eta )]^{D+1}\langle 0_{\mathrm{R}}|T_{i}^{k}%
\left[ g_{lm}^{\mathrm{R}},\varphi _{\mathrm{R}}\right] |0_{\mathrm{R}%
}\rangle +\langle T_{i}^{k}\left[ g_{lm}^{\prime },\varphi \right] \rangle ^{%
\mathrm{(an)}},  \label{conftransemt}
\end{equation}%
where the second term on the right is determined by the trace anomaly. In
odd spacetime dimensions the conformal anomaly is absent and the
corresponding part vanishes: $\langle T_{i}^{k}\left[ g_{lm}^{\prime
},\varphi \right] \rangle ^{\mathrm{(an)}}=0$ for even $D$. The vacuum
expectation value of the energy-momentum tensor in coordinates (\ref{dsRW})
is obtained by the standard coordinate transformation formulae. For a second
rank tensor $A_{ik}$, which is diagonal in coordinates $x^{\prime i}=(\eta
,\xi ,\mathbf{x}^{\prime })$, the transformation to coordinates $x^{i}=(\eta
,r,\theta ,\theta _{2},\ldots ,\theta _{D})$ has the form
\begin{eqnarray}
A_{0}^{0} &=&A_{0}^{\prime 0},\;A_{1}^{1}=A_{1}^{\prime 1}+\Omega ^{2}\sin
^{2}\theta (A_{2}^{\prime 2}-A_{1}^{\prime 1}),  \notag \\
A_{1}^{2} &=&\Omega ^{2}\sin \theta \frac{\cos \theta -r\gamma }{r}%
(A_{2}^{\prime 2}-A_{1}^{\prime 1}),  \label{A21} \\
A_{2}^{2} &=&A_{2}^{\prime 2}+\Omega ^{2}\sin ^{2}\theta (A_{1}^{\prime
1}-A_{2}^{\prime 2}),\;A_{l}^{l}=A_{2}^{\prime 2},\;l=3,\ldots ,D.  \notag
\end{eqnarray}

In this paper, as a Rindler counterpart we shall take the vacuum
energy--momentum tensor induced by an infinite plate moving by uniform
proper acceleration through the Fulling--Rindler vacuum. We shall assume
that the plate is located in the right Rindler wedge and has the coordinate $%
\xi =b$. In coordinates $x^{i}$ the boundary $\xi =b$ is presented by the
hypersurface%
\begin{equation}
\sqrt{1+r^{2}}-r\cos \theta =1/b_{0},\;b_{0}=b/\xi _{0}\text{.}
\label{BoundEq}
\end{equation}%
The corresponding normal has the components%
\begin{equation}
n^{l}=\frac{b_{0}}{ra(\eta )}(0,\sqrt{1+r^{2}}(1-\sqrt{1+r^{2}}/b_{0}),-\sin
\theta ,0,\ldots ,0).  \label{nl}
\end{equation}%
We consider the cases of scalar and electromagnetic fields separately.

\section{Vacuum expectation values for the energy-momentum tensor: Scalar
field}

\label{sec:Scalar}

In this section we consider a conformally coupled massless scalar field $%
\varphi (x)$ on background of spacetime with the line element (\ref{dsRW}).
The corresponding field equation has the form
\begin{equation}
\left( \nabla _{l}\nabla ^{l}+\frac{D-1}{4D}R\right) \varphi (x)=0,
\label{fieldeq}
\end{equation}%
where $R$ is the Ricci scalar for the RW spacetime. We assume that
the field satisfies the Robin boundary condition
\begin{equation}
(A+Bn^{l}\nabla _{l})\varphi (x)=0,  \label{BC}
\end{equation}%
on the hypersurface (\ref{BoundEq}).

The expectation value of the energy-momentum tensor induced by the presence
of an infinite plane boundary moving with uniform acceleration through the
Fulling--Rindler vacuum was investigated in \cite{CandD,Saha02}. For a
scalar field $\varphi _{\mathrm{R}}(x^{\prime })$ it is presented in the
decomposed form:
\begin{equation}
\langle 0_{\mathrm{R}}|T_{i}^{k}[g_{lm}^{\mathrm{R}},\varphi _{\mathrm{R}%
}]|0_{\mathrm{R}}\rangle =\langle \tilde{0}_{\mathrm{R}}|T_{i}^{k}[g_{lm}^{%
\mathrm{R}},\varphi _{\mathrm{R}}]|\tilde{0}_{\mathrm{R}}\rangle +\langle T_{%
\mathrm{(R)}i}^{k}\rangle ^{\mathrm{(b)}}.  \label{TikR}
\end{equation}%
In this formula, $|0_{\mathrm{R}}\rangle $ are $|\tilde{0}_{\mathrm{R}%
}\rangle $ are the vacuum states for the Rindler spacetime in presence and
absence of the plate respectively \ and $\langle T_{\mathrm{(R)}%
i}^{k}\rangle ^{\mathrm{(b)}}$ is the part of the vacuum energy-momentum
tensor induced by the plate. For the part without boundaries one has
\begin{equation}
\langle \tilde{0}_{\mathrm{R}}|T_{i}^{k}[g_{lm}^{\mathrm{R}},\varphi _{%
\mathrm{R}}]|\tilde{0}_{\mathrm{R}}\rangle =\frac{a_{D}\xi ^{-D-1}}{%
2^{D-1}\pi ^{D/2}\Gamma (D/2)}\mathrm{diag}\left( -1,1/D,\ldots ,1/D\right) ,
\label{TikR0}
\end{equation}%
with the notation
\begin{equation}
a_{D}=\int_{0}^{\infty }\frac{\omega ^{D}d\omega }{e^{2\pi \omega }+(-1)^{D}}%
\prod_{l=1}^{l_{m}}\left[ \left( \frac{D-1-2l}{2\omega }\right) ^{2}+1\right]
,  \label{aD}
\end{equation}%
where $l_{m}=D/2-1$ for even $D>2$ and $l_{m}=(D-1)/2$ for odd $D>1$, and
the value for the product over $l$ is equal to 1 for $D=1,2,3$.

For a scalar field $\varphi _{\mathrm{R}}(x^{\prime })$ satisfying the Robin
boundary condition
\begin{equation}
\left( A_{\mathrm{R}}+B_{\mathrm{R}}n^{\prime }{}_{\mathrm{R}}^{l}\nabla
_{l}^{\prime }\right) \varphi _{\mathrm{R}}(x^{\prime })=0,\quad \xi
=b,\quad n^{\prime }{}_{\mathrm{R}}^{l}=\delta _{1}^{l},  \label{boundRind}
\end{equation}%
with constant coefficients $A_{\mathrm{R}}$ and $B_{\mathrm{R}}$, the
boundary induced part in the region $\xi >a$ is given by the formula \cite%
{Saha02}
\begin{equation}
\langle T_{\mathrm{(R)}i}^{k}\rangle ^{\mathrm{(b)}}=\frac{-2^{1-D}\delta
_{i}^{k}b^{-D-1}}{\pi ^{(D+1)/2}D\Gamma \left( \frac{D-1}{2}\right) }%
\int_{0}^{\infty }dx\,x^{D}\int_{0}^{\infty }d\omega \frac{\bar{I}_{\omega
}(x)}{\bar{K}_{\omega }(x)}F^{(i)}[K_{\omega }(x\xi /b)].  \label{TikRb}
\end{equation}%
Here the functions $F^{(i)}[g(z)]$ have the form
\begin{eqnarray}
F^{(0)}[g(z)] &=&g^{\prime 2}(z)+\frac{D-1}{z}g(z)g^{\prime }(z)+\left[
1-(2D-1)\frac{\omega ^{2}}{z^{2}}\right] g^{2}(z),  \notag \\
F^{(1)}[g(z)] &=&-Dg^{\prime 2}(z)-\frac{D-1}{z}g(z)g^{\prime }(z)+D\left( 1+%
\frac{\omega ^{2}}{z^{2}}\right) g^{2}(z),  \label{Figz} \\
F^{(i)}[g(z)] &=&g^{\prime 2}(z)+\left( \frac{\omega ^{2}}{z^{2}}-\frac{D+1}{%
D-1}\right) g^{2}(z),\quad i=2,\ldots ,D.  \notag
\end{eqnarray}%
In Eq. (\ref{TikRb}), $I_{\omega }(z)$ and $K_{\omega }(z)$ are the modified
Bessel functions and for a given function $f(z)$ we use the notation
\begin{equation}
\bar{f}(z)=A_{\mathrm{R}}f(z)+(B_{\mathrm{R}}/b)zf^{\prime }(z).
\label{barnot}
\end{equation}%
The expression for the boundary part of the vacuum energy-momentum tensor in
the region $\xi <a$ is obtained from formula (\ref{TikRb}) by the
replacements $I_{\omega }\rightleftarrows K_{\omega }$.

The formulae given above allow us to present the RW vacuum expectation value
in coordinates $x^{\prime i}$\ in the form similar to (\ref{TikR}):
\begin{equation}
\langle 0_{\mathrm{RW}}|T_{i}^{k}\left[ g_{lm}^{\prime },\varphi \right] |0_{%
\mathrm{RW}}\rangle =\langle \tilde{0}_{\mathrm{RW}}|T_{i}^{k}\left[
g_{lm}^{\prime },\varphi \right] |\tilde{0}_{\mathrm{RW}}\rangle +\langle
T_{i}^{k}\left[ g_{lm}^{\prime },\varphi \right] \rangle ^{\mathrm{(b)}},
\label{TikdS}
\end{equation}%
where $\langle \tilde{0}_{\mathrm{RW}}|T_{i}^{k}\left[ g_{lm}^{\prime
},\varphi \right] |\tilde{0}_{\mathrm{RW}}\rangle $ is the vacuum
expectation value in the RW spacetime without boundaries and the part $%
\langle T_{i}^{k}\left[ g_{lm}^{\prime },\varphi \right] \rangle ^{\mathrm{%
(b)}}$ is induced by the boundary (\ref{BoundEq}). Conformally transforming
the Rindler results one finds
\begin{eqnarray}
\langle \tilde{0}_{\mathrm{RW}}|T_{i}^{k}\left[ g_{lm}^{\prime },\varphi %
\right] |\tilde{0}_{\mathrm{RW}}\rangle  &=&[\xi /a(\eta )]^{D+1}\langle
\tilde{0}_{\mathrm{R}}|T_{i}^{k}[g_{lm}^{\mathrm{R}},\varphi _{\mathrm{R}}]|%
\tilde{0}_{\mathrm{R}}\rangle +\langle T_{i}^{k}\left[ g_{lm}^{\prime
},\varphi \right] \rangle ^{\mathrm{(an)}},  \label{TikdS0} \\
\langle T_{i}^{k}\left[ g_{lm}^{\prime },\varphi \right] \rangle ^{\mathrm{%
(b)}} &=&[\xi /a(\eta )]^{D+1}\langle T_{\mathrm{(R)}i}^{k}\rangle ^{\mathrm{%
(b)}}.  \label{TikdSb}
\end{eqnarray}%
Under the conformal transformation $g_{ik}^{\prime }=[a(\eta )/\xi
]^{2}g_{ik}^{\mathrm{R}}$, the field $\varphi _{\mathrm{R}}$ is changed by
the rule
\begin{equation}
\varphi (x^{\prime })=[\xi /a(\eta )]^{(D-1)/2}\varphi _{\mathrm{R}%
}(x^{\prime }).  \label{FieldTrans}
\end{equation}%
Now by comparing boundary conditions (\ref{BC}), (\ref{boundRind}) and
taking into account Eq. (\ref{FieldTrans}), one obtains the relation between
the coefficients in the boundary conditions:
\begin{equation}
a(\eta )A/B=bA_{\mathrm{R}}/B_{\mathrm{R}}+(1-D)/2.  \label{relcoef}
\end{equation}%
As it is seen from this relation, the Dirichlet boundary condition in the
problem on the RW bulk ($B=0$) corresponds to the Dirichlet boundary
condition in the conformally related problem for the Rindler spacetime. For
the case of the Neumann boundary condition in the RW bulk ($A=0$) the
corresponding problem in the Rindler spacetime is of the Robin type with $%
bA_{\mathrm{R}}/B_{\mathrm{R}}=(D-1)/2$.

As before, we shall present the corresponding components in coordinates $%
x^{i}$ in the form of the sum of purely RW and boundary parts:
\begin{equation}
\langle 0_{\mathrm{RW}}|T_{i}^{k}\left[ g_{lm},\varphi \right] |0_{\mathrm{RW%
}}\rangle =\langle \tilde{0}_{\mathrm{RW}}|T_{i}^{k}\left[ g_{lm},\varphi %
\right] |\tilde{0}_{\mathrm{RW}}\rangle +\langle T_{i}^{k}\rangle ^{\mathrm{%
(b)}}.  \label{TikdS1}
\end{equation}%
By using the relations (\ref{A21}) for the purely RW part one finds (for the
vacuum polarization in RW spacetimes see \cite{Birrell,Grib94,Bord97} and
references therein)
\begin{equation}
\langle \tilde{0}_{\mathrm{RW}}|T_{i}^{k}\left[ g_{lm},\varphi \right] |%
\tilde{0}_{\mathrm{RW}}\rangle =\frac{2a_{D}[a(\eta )]^{-D-1}}{(4\pi
)^{D/2}\Gamma (D/2)}\mathrm{diag}\left( -1,1/D,\ldots ,1/D\right) +\langle
T_{i}^{k}\left[ g_{lm},\varphi \right] \rangle ^{\mathrm{(an)}}.
\label{Tik0dSst}
\end{equation}%
In particular, for $D=3$ we have \cite{Birrell}%
\begin{equation}
\langle T_{i}^{k}\left[ g_{lm},\varphi \right] \rangle ^{\mathrm{(an)}}=%
\frac{^{(3)}H_{i}^{k}-^{(1)}H_{i}^{k}/6}{2880\pi ^{2}},  \label{Tikan}
\end{equation}%
where the expressions for the tensors $^{(j)}H_{i}^{k}$ are given in \cite%
{Birrell}. Now it can be easily checked that for the static case, $a(\eta )=%
\mathrm{const}$, one has $\langle \tilde{0}_{\mathrm{RW}}|T_{i}^{k}\left[
g_{lm},\varphi \right] |\tilde{0}_{\mathrm{RW}}\rangle =0$. In the special
case of the power-law expansion, $a(t)=\alpha t^{c}$, with $t$ being the
synchronous time coordinate, we find%
\begin{equation}
\langle \tilde{0}_{\mathrm{RW}}|T_{i}^{k}\left[ g_{lm},\varphi \right] |%
\tilde{0}_{\mathrm{RW}}\rangle =\frac{c(c^{2}-6c+3)\left( 3c-4\right) }{%
2880\pi ^{2}t^{4}}\mathrm{diag}(\frac{3c}{3c-4},1,1,1).  \label{TikRWD3}
\end{equation}%
The corresponding energy density is negative for $|c-3|<\sqrt{6}$.

\begin{figure}[tbph]
\begin{center}
\epsfig{figure=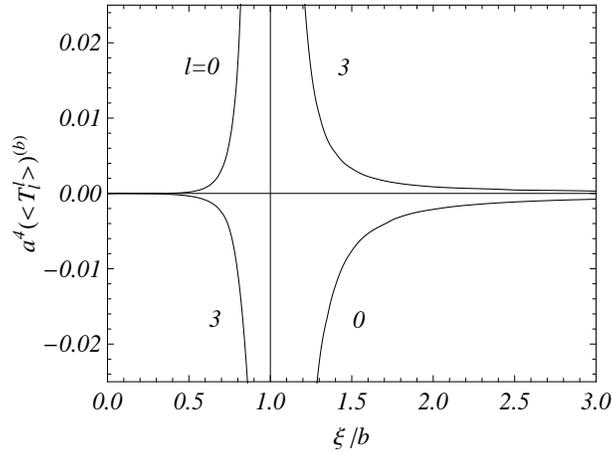,width=8cm,height=6.cm}
\end{center}
\caption{Boundary induced parts in the vacuum energy density and ${}_{3}^{3}$%
-stress for $D=3$ Dirichlet scalar field.}
\label{fig1}
\end{figure}

For the boundary induced energy-momentum tensor in coordinates $x^{\prime i}$
the spatial part is not isotropic and the corresponding part in coordinates $%
x^{i}$ is more complicated (no summation over $l$):
\begin{eqnarray}
\langle T_{l}^{l}\rangle ^{\mathrm{(b)}} &=&[\xi /a(\eta )]^{D+1}\langle T_{%
\mathrm{(R)}l}^{l}\rangle ^{\mathrm{(b)}},\;l=0,3,\ldots ,D,  \notag \\
\langle T_{l}^{l}\rangle ^{\mathrm{(b)}} &=&[\xi /a(\eta )]^{D+1}\left[
\langle T_{\mathrm{(R)}l}^{l}\rangle ^{\mathrm{(b)}}+(-1)^{l}\Omega ^{2}\sin
^{2}\theta (\langle T_{\mathrm{(R)}1}^{1}\rangle ^{\mathrm{(b)}}-\langle T_{%
\mathrm{(R)}2}^{2}\rangle ^{\mathrm{(b)}})\right] ,\;l=1,2,
\label{Tik0dSstb} \\
\langle T_{1}^{2}\rangle ^{\mathrm{(b)}} &=&[\xi /a(\eta )]^{D+1}\Omega
^{2}\sin \theta \frac{\cos \theta -r\gamma }{r}(\langle T_{\mathrm{(R)}%
2}^{2}\rangle ^{\mathrm{(b)}}-\langle T_{\mathrm{(R)}1}^{1}\rangle ^{\mathrm{%
(b)}}).  \notag
\end{eqnarray}%
As we see the resulting energy-momentum tensor is non-diagonal. Note that
for the case of the power-law expansion the ratio of the boundary induced
and boundary free parts at a given spatial point behaves as $t^{4(1-c)}$ in
the model with $D=3$. Hence, at early stages of the cosmological expansion
the boundary induced part dominates for $c>1$. In figure \ref{fig1} we have
plotted the boundary induced parts in the vacuum energy density ($l=0$) and $%
_{3}^{3}$-stress ($l=3$) as functions of the ratio $\xi /b$ for $D=3$ scalar
field with Dirichlet boundary condition. The corresponding energy density is
positive in the region $\xi <b$ and negative for $\xi >b$. In the case of
Robin condition the energy density can be either negative or positive in
dependence of the coefficient in the boundary condition.

\section{Electromagnetic field}

\label{sec:Elmag}

The electromagnetic field is conformally invariant in $D=3$. The vacuum
expectation value of the energy-momentum tensor induced by the presence of
conducting plate moving with uniform acceleration through the
Fulling--Rindler vacuum is investigated in Refs. \cite{CandD,Saha02}. We
will assume that the plate is a perfect conductor with the standard boundary
conditions of vanishing of the normal component of the magnetic field and
the tangential components of the electric field, evaluated at the local
inertial frame in which the conductor is instantaneously at rest. As in the
case of a scalar field, the expectation value of the energy-momentum tensor
is presented in the form (\ref{TikR}), where the boundary free part is given
by the formula%
\begin{equation}
\langle \tilde{0}_{\mathrm{R}}|T_{i}^{k}[g_{lm}^{\mathrm{R}},\varphi _{%
\mathrm{R}}]|\tilde{0}_{\mathrm{R}}\rangle =\frac{11}{240\pi ^{2}\xi ^{4}}%
\mathrm{diag}\left( -1,1/3,1/3,1/3\right) .  \label{TikREl}
\end{equation}%
In the region $\xi >b$ for the boundary induced part one has the expression
\cite{CandD,Saha02}
\begin{equation}
\langle T_{\mathrm{(R)}i}^{k}\rangle ^{\mathrm{(b)}}=-\frac{\delta _{i}^{k}}{%
4\pi ^{2}b^{4}}\int_{0}^{\infty }dx\,x^{3}\int_{0}^{\infty }d\omega \left[
\frac{I_{\omega }(x)}{K_{\omega }(x)}+\frac{I_{\omega }^{\prime }(x)}{%
K_{\omega }^{\prime }(x)}\right] F_{\mathrm{em}}^{(i)}[K_{\omega }(x\xi /b)],
\label{TikbEl}
\end{equation}%
with the notations%
\begin{eqnarray}
F_{\mathrm{em}}^{(i)}[g(z)] &=&(-1)^{i}g^{\prime 2}(z)+[1-(-1)^{i}\omega
^{2}/z^{2}]g^{2}(z),\;i=0,1,  \notag \\
F_{\mathrm{em}}^{(2)}[g(z)] &=&F_{\mathrm{em}}^{(3)}[g(z)]=-g^{2}(z).
\label{Fiem}
\end{eqnarray}%
The corresponding formula in the region $\xi <b$ is obtained from (\ref%
{TikbEl}) by the replacements $I_{\omega }\rightleftarrows K_{\omega }$. By
taking into account that $[I_{\omega }(x)K_{\omega }(x)]^{\prime }<0$, we
see that $\langle T_{\mathrm{(R)}2}^{2}\rangle ^{\mathrm{(b)}}>0$ for $\xi
>b $ and $\langle T_{\mathrm{(R)}2}^{2}\rangle ^{\mathrm{(b)}}<0$ for $\xi
<b $. For the perpendicular stress one has $\langle T_{\mathrm{(R)}%
1}^{1}\rangle ^{\mathrm{(b)}}>0$ in both regions. The boundary induced
energy density is positive/negative in the region $\xi <b$/$\xi >b$.

The vacuum expectation value of the energy-momentum tensor in the RW bulk is
presented in the form (\ref{TikdS1}), where the boundary free part is given
by the expression%
\begin{equation}
\langle \tilde{0}_{\mathrm{RW}}|T_{i}^{k}\left[ g_{lm},\varphi \right] |%
\tilde{0}_{\mathrm{RW}}\rangle =\frac{11a^{-4}(\eta )}{240\pi ^{2}}\mathrm{%
diag}\left( -1,1/3,1/3,1/3\right) +\frac{62^{(3)}H_{i}^{k}+3^{(1)}H_{i}^{k}}{%
2880\pi ^{2}}.  \label{TikElfree}
\end{equation}%
As in the case for a scalar field, this expectation value vanishes for the
static RW spacetime. For the power-law expansion, $a(t)=\alpha t^{c}$, from (%
\ref{TikElfree}) we find%
\begin{eqnarray}
\langle \tilde{0}_{\mathrm{RW}}|T_{i}^{k}\left[ g_{lm},\varphi \right] |%
\tilde{0}_{\mathrm{RW}}\rangle &=&\frac{c(31c^{2}+54c-27)(3c-4)}{1440\pi
^{2}t^{4}}\mathrm{diag}(\frac{3c}{3c-4},1,1,1)  \notag \\
&&-\frac{c(c-2)}{18\pi ^{2}a^{2}(t)t^{2}}\mathrm{diag}(\frac{3c}{c-2},1,1,1).
\label{TikRW3DEl}
\end{eqnarray}%
For $c<1$ the second term on the right of this formula dominates at late
stages of the cosmological expansion and the corresponding energy density is
negative.

For the boundary induced part we have formulae (\ref{Tik0dSstb}) with $D=3$
and with $\langle T_{\mathrm{(R)}i}^{k}\rangle ^{\mathrm{(b)}}$ given by (%
\ref{TikbEl}) for the region $\xi >b$. For points near the boundary the
leading terms in the asymptotic expansions for the components of the
energy-momentum tensor have the form%
\begin{eqnarray}
\langle T_{0}^{0}\rangle ^{\mathrm{(b)}} &\approx &\frac{-2\langle
T_{1}^{1}\rangle ^{\mathrm{(b)}}}{b_{0}^{2}\sin ^{2}\theta }\approx \frac{%
2\langle T_{2}^{2}\rangle ^{\mathrm{(b)}}}{b_{0}^{2}\sin ^{2}\theta -1}%
\approx -2\langle T_{3}^{3}\rangle ^{\mathrm{(b)}}\approx \frac{(1-\xi
/b)^{-3}}{30\pi ^{2}a^{4}(\eta )},  \notag \\
\langle T_{1}^{2}\rangle ^{\mathrm{(b)}} &\approx &\frac{r\gamma -\cos
\theta }{r}\frac{a^{-4}(\eta )b_{0}^{2}\sin \theta }{60\pi ^{2}(1-\xi /b)^{3}%
}.  \label{TikElAs}
\end{eqnarray}%
In the asymptotic term for the off-diagonal component $r$ and $\theta $ are
related by (\ref{BoundEq}). Near the boundary the total energy-momentum
tensor is dominated by the boundary-induced part.

In the limit $\xi \rightarrow 0$ we have the following asymptotic formulae
(no summation over $l$)%
\begin{eqnarray}
\langle T_{0}^{0}\rangle ^{\mathrm{(b)}} &\approx &\langle T_{1}^{1}\rangle
^{\mathrm{(b)}}\approx -\;\langle T_{l}^{l}\rangle ^{\mathrm{(b)}}\approx
\frac{-1.326(\xi /b)^{4}}{8\pi ^{2}a^{4}(\eta )\ln (2b/\xi )},\;l=2,3,
\notag \\
\langle T_{1}^{2}\rangle ^{\mathrm{(b)}} &\approx &-b _{0}^{2}\sin
\theta \frac{1.326(\xi /b)^{6}\sin ^{2}(\theta /2)}{2\pi
^{2}a^{4}(\eta )r\ln (2b/\xi )}.  \label{TikElsmall}
\end{eqnarray}%
This limit corresponds to large values of the coordinate $r$ with the
relation $\xi /\xi _{0}\approx \lbrack 2r\sin ^{2}(\theta /2)]^{-1}$. Now we
turn to the limit $\xi /b\rightarrow \infty $. In terms of the coordinates $r
$ and $\theta $, this limit corresponds to large values of $r$ and small
values of $\theta $ with $\xi /\xi _{0}\approx 2r/\left( r^{2}\theta
^{2}+1\right) $. To the leading order for the boundary induced part we have%
\begin{equation}
\langle T_{i}^{k}\rangle ^{\mathrm{(b)}}=\frac{a^{-4}(\eta )}{96\ln ^{2}(\xi
/b)}\mathrm{diag}\left( 1,-1/3,-1/3,-1/3\right) .  \label{TikEllarge}
\end{equation}%
In this limit the total energy-momentum tensor is dominated by the boundary
free part (\ref{TikElfree}).

\section{Conclusion}

\label{sec:Conc}

In the investigations of the Casimir effect the calculation of the
local densities of the vacuum characteristics is of special
interest. In particular, these include the vacuum expectation
value of the energy--momentum tensor. In addition to describing
the physical structure of the quantum field at a given point, the
energy--momentum tensor acts as the source of gravity in the
Einstein equations. It therefore plays an important role in
modelling a self-consistent dynamics involving the gravitational
field.

In the present paper we have investigated the vacuum expectation
value of the energy-momentum tensor for scalar and electromagnetic
fields induced by the boundary, defined by Eq. (\ref{BoundEq}), on
background of RW spacetime with negative spatial curvature. For a
scalar field the Robin boundary condition is imposed and for the
electromagnetic field we have assumed that the boundary is a
perfect conductor. In order to obtain the vacuum expectation
values we have used the corresponding results for a plate moving
with constant proper acceleration through the Fulling--Rindler
vacuum and the conformal relation between the $k=-1$ RW and
Rindler spacetimes.

For the general case of the scale factor we have presented the
vacuum energy-momentum tensor as the sum of the boundary free and
boundary induced
parts. The boundary free parts are given by the standard formulae (\ref%
{Tik0dSst}), (\ref{Tikan}) for a scalar field and by (\ref{TikElfree}) for
the electromagnetic field. In the special case of the power-law scale factor
the corresponding expressions take the forms (\ref{TikRWD3}) and (\ref%
{TikRW3DEl}), respectively. The boundary induced part in the vacuum
expectation value of the energy-momentum tensor is non-diagonal and is given
by the expressions (\ref{Tik0dSstb}) with $\langle T_{\mathrm{(R)}%
i}^{k}\rangle ^{\mathrm{(b)}}$ defined by formulae (\ref{TikRb}) and (\ref%
{TikbEl}) for scalar and electromagnetic fields in the region $\xi >b$. The
corresponding formulae for the region $\xi <b$ are obtained by the
replacements $I_{\omega }\rightleftarrows K_{\omega }$. In the case of the
electromagnetic field the boundary induced energy density is positive
(negative) in the region $\xi <b$ ($\xi >b$). For the power-law expansion
with $a(t)\propto t^{c}$, $c>1$, at a given spatial point the ratio of the
boundary induced and boundary free parts in the vacuum energy-momentum
tensor behaves as $t^{4(1-c)}$ in the model with $D=3$ and at early stages
of the cosmological expansion the boundary induced part dominates.

\section*{Acknowledgments}

A.A.S. was supported by the Armenian Ministry of Education and Science Grant
No. 119.

\end{document}